\documentclass[12pt]{article}

\usepackage{scicite}

\usepackage{times}

\usepackage{amsmath}
\usepackage{amsfonts}
\usepackage{amssymb}
\usepackage{graphicx}% Include figure files
\usepackage{dcolumn}% Align table columns on decimal point
\usepackage{xfrac}
\usepackage{color}% text color
\usepackage{hyperref}% add hypertext capabilities
\usepackage{tabularx}

\newcommand{\CBI}{CrCl$_{3-x-y}$Br$_x$I$_y$}
\newcommand{\CCC}{CrCl$_{0.8}$Br$_{1.2}$I$_{1.0}$}
\newcommand{\CBmax}{CrCl$_{2.55}$Br$_{0.45}$}
\newcommand{\ClBr}{CrCl$_{3-x}$Br$_x$}
\newcommand{\BrI}{CrBr$_{y}$I$_{3-y}$}
\newcommand{\ClI}{CrCl$_{x}$I$_{3-x}$}
\newcommand{\CC}{CrCl$_3$}
\newcommand{\CB}{CrBr$_3$}
\newcommand{\CI}{CrI$_3$}
\newcommand{\CGT}{Cr$_2$Ge$_2$Te$_6$}
\newcommand{\FGT}{Fe$_3$GeTe$_2$}
\newcommand{\C}{$^\circ$C}
\newenvironment{sciabstract}{%
\begin{quote} \bf}
{\end{quote}}

\title{Accessing new magnetic regimes by tuning the ligand spin-orbit coupling in van der Waals magnets}

\author
{Thomas~A.~Tartaglia,$^1$
Joseph~N.~Tang,$^1$
Jose~L.~Lado,$^2$\\
Faranak~Bahrami,$^1$
Mykola~Abramchuk,$^1$
Gregory~T.~McCandless,$^3$\\
Meaghan~C.~Doyle,$^1$
Kenneth~S.~Burch,$^1$
Ying~Ran,$^1$\\
Julia~Y.~Chan,$^3$
Fazel~Tafti,$^{1\dag}$\\
\normalsize{${}^{1}$Department of Physics, Boston College, Chestnut Hill, MA 02467, USA}\\
\normalsize{${}^{2}$Department of Applied Physics, Aalto University, Espoo, Finland}\\
\normalsize{${}^{3}$Department of Chemistry and Biochemistry, University of Texas at Dallas, Richardson, TX 75080, USA}\\
\normalsize{$^\dag$fazel.tafti@bc.edu}
}

\date{}

\begin{document}


\baselineskip24pt


\maketitle

% The abstract should be a single paragraph, not to exceed 250 words and ideally closer to 200, written in plain language that a general reader can understand. It should include
% An opening sentence that states the question/problem addressed by the research AND
% Enough background content to give context to the study AND
% A brief statement of primary results AND
% A short concluding sentence.
% Do not include citations or undefined abbreviations in the abstract. Any abbreviations that appear in the title should be defined in the abstract.

\noindent\textbf{One Sentence Summary:} New magnetic regimes and phenomena are revealed by tuning the competition between two types of spin-orbit coupling.

\pagebreak
\begin{sciabstract}
Van der Waals (VdW) materials have opened new directions in the study of low dimensional magnetism.
A largely unexplored arena is the intrinsic tuning of VdW magnets toward new ground-states.
The chromium trihalides provided the first such example with a change of inter-layer magnetic coupling emerging upon exfoliation.
Here, we take a different approach to engineer new ground-states, not by exfoliation, but by tuning the spin-orbit coupling (SOC) of the non-magnetic ligand atoms (Cl,Br,I).
We synthesize a three-halide series, \CBI, and map their magnetic properties as a function of Cl, Br, and I content.
The resulting triangular phase diagrams unveil a frustrated regime near \CC.
First-principles calculations confirm that the frustration is driven by a competition between the chromium and halide SOCs.
Furthermore, we reveal a field-induced change of inter-layer coupling in the bulk of \CBI\ crystals at the same field as in the exfoliation experiments.

\end{sciabstract}

\pagebreak

% In setting up this template for *Science Advances* papers, both
% the \section* command and the \paragraph* command are used for topical
% divisions.  Which you use will of course depend on the type of paper
% you're writing.  Review Articles tend to have displayed headings, for
% which \section* is more appropriate; Research Articles, when they have
% formal topical divisions at all, tend to signal them with bold text
% that runs into the paragraph, for which \paragraph* is the right
% choice.  Either way, use the asterisk (*) modifier, as shown, to

\section*{Introduction}
% The manuscript should start with a brief introduction that lays out the problem addressed by the research and describes the paper's importance. The scientific question being investigated should be described in detail. The introduction should provide sufficient background information to make the article understandable to readers in other disciplines, and provide enough context to ensure that the implications of the experimental findings are clear.

%%
Spin-orbit coupling (SOC) is an essential ingredient of exotic magnetic phenomena.
For example, the Dzyaloshinski-Moriya interaction that leads to the formation of Skyrmions originates from a combination of SOC and broken inversion symmetry~\cite{luo_strong_2019}.
The Kitaev interaction that leads to long-range entanglement in quantum spin liquids is also rooted in SOC~\cite{takagi_concept_2019,bahrami_thermodynamic_2019}.
Ferromagnetic (FM) ordering in single atomic layers of 2D magnets is a result of SOC and magnetic anisotropy according to the Mermin-Wagner theorem~\cite{huang_layer-dependent_2017,jiang_controlling_2018,gong_discovery_2017,mermin_absence_1966,burch_magnetism_2018}.
Despite the fundamental importance of SOC, little has been done to tune this interaction beyond the single-ion level.
In this manuscript, we generate the entire magnetic phase diagram of chromium trihalides by tuning the ligand SOC.
This is done by varying the ratio of the non-magnetic ligand atoms (Cl,Br,I) in \CBI\ without affecting the magnetic atom (Cr).

Our experiments were motivated by the following model Hamiltonian recently proposed for chromium trihalides~\cite{lado_origin_2017,torelli_calculating_2018,xu_interplay_2018,lee_fundamental_2020}.
\begin{equation}
\label{eq:HCOMP}
\mathcal{H}= -J\sum_{\langle i,j \rangle}\mathbf{S_i\cdot S_j} -D\sum_i(S_i^z)^2 -\lambda\sum_{\langle i,j \rangle}S_i^z S_j^z + H_{\textrm{frust.}}
\end{equation}
It describes the Cr$^{3+}$ ions with $S=3/2$ and isotropic Heisenberg coupling ($J$) on a honeycomb lattice (Fig.~\ref{fig:CBI}).
\textcolor{black}{Magnetic anisotropy is controlled by the single-ion anisotropy ($D$) and the anisotropic exchange ($\lambda$).
The main contribution to $D$ is from the SOC of Cr.
The main contribution to $\lambda$ is from the SOC of the heavy ligands (Br and I), but Cr and Cl could also contribute to a lesser extent.}
Other interactions that promote magnetic frustration, such as the Kitaev or dipolar interactions, are contained in $H_{\text{frust.}}$.
The in-plane anisotropy in \CC~\cite{cable_neutron_1961} implies $D<0$ and the out-of-plane anisotropy in \CI~\cite{mcguire_coupling_2015} implies $\lambda>0$.
Thus, a competition is built into Eq.~\ref{eq:HCOMP} that could drive the system into a frustrated regime due to $H_{\text{frust.}}$ when $D$ and $\lambda$ acquire comparable magnitudes (with opposite signs).
The atomic SOC values in Cr, Cl, Br, and I are 90, 40, 220, and 580~meV, respectively~\cite{martin_table_1971}.
From here, we expect to see the effect of $H_{\text{frust.}}$ near \CC\ where $D$ and $\lambda$ could cancel each other, but not in \CB\ or \CI\ where a strong $\lambda$ leads to a dominant out-of-plane FM order.
To search for such a frustrated regime, we grew a series of three-halide crystals \CBI\ and mapped their Curie temperature ($T_C$), Weiss temperature ($\Theta_W$), and frustration index ($f=\Theta_W/T_C$), as a function of Cl, Br, and I content.
As expected, our experiments revealed a frustrated regime near \CC\ with maximum $f$ in \CBmax.

In addition to finding a frustrated regime, we reveal a field-induced change of inter-layer coupling from antiferromagnetic (AF) to ferromagnetic (FM) in the bulk \CBI\ crystals.
So far, such metamagnetic transition has been observed only in exfoliation experiments due to the monoclinic stacking of two atomic layers (bilayer) of \CI~\cite{huang_layer-dependent_2017,jiang_controlling_2018,li_pressure-controlled_2019} and \CB~\cite{chen_direct_2019}.
We show, for the first time, that a similar metamagnetic transition can be engineered in the bulk of \CBI\ intrinsically.

%%%%%%%%%%%%%%%%%%%%%%%%%%%%%%%%%%%%%%%%%%%%%%%%%%%%%%%%%%%%%%%%%%%%%%%%%%%%%%%%%%%%%%%%%%%%%%%%%%%%%%%%%%%%%%%%%%%%%%%%%%%
%%%%%%%%%%%%%%%%%%%%%%%%%%%%%%%%%%%%%%%%%%%%%%%%%%%%%%%%% SECTIONS %%%%%%%%%%%%%%%%%%%%%%%%%%%%%%%%%%%%%%%%%%%%%%%%%%%%%%%%
%%%%%%%%%%%%%%%%%%%%%%%%%%%%%%%%%%%%%%%%%%%%%%%%%%%%%%%%%%%%%%%%%%%%%%%%%%%%%%%%%%%%%%%%%%%%%%%%%%%%%%%%%%%%%%%%%%%%%%%%%%%

\section*{Results}
\subsection*{VdW Alloys}
%%%%%%%%%%%%%%%%%%%%%%%%%%%%%%%%%%%%%%%%%%%%%%%%%%%
%%%%%%%%%%%%%%%%%%%%% FIGURE 1%%%%%%%%%%%%%%%%%%%%%
\begin{figure}%[htb!]
\centering
\includegraphics[width=0.45\textwidth]{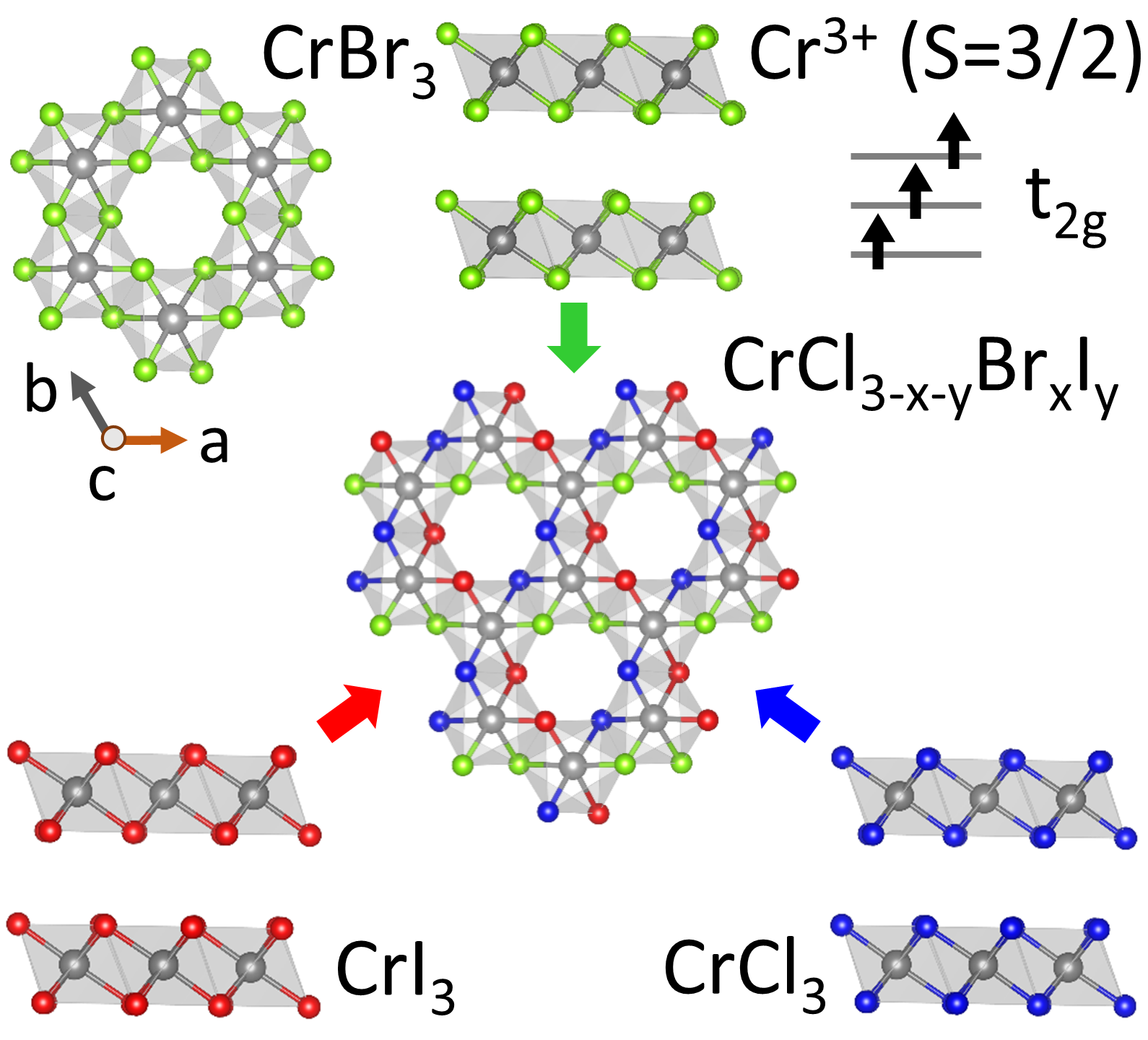}
\caption{\label{fig:CBI}
The \CBI\ alloys with a honeycomb layered structure were synthesized by a CVT process from mixtures of \CC, \CB, and \CI\ in appropriate ratios.
The gray, blue, green, and red spheres represent Cr$^{3+}$, Cl$^-$, Br$^-$, and I$^-$, respectively.
Cr$^{3+}$ is a spin 3/2 ion with three electrons in the $t_{\mathrm{2g}}$ manifold.
}
\rule{\textwidth}{0.5pt}
\end{figure}
\begin{figure}%[htb!]
\centering
\includegraphics[width=0.45\textwidth]{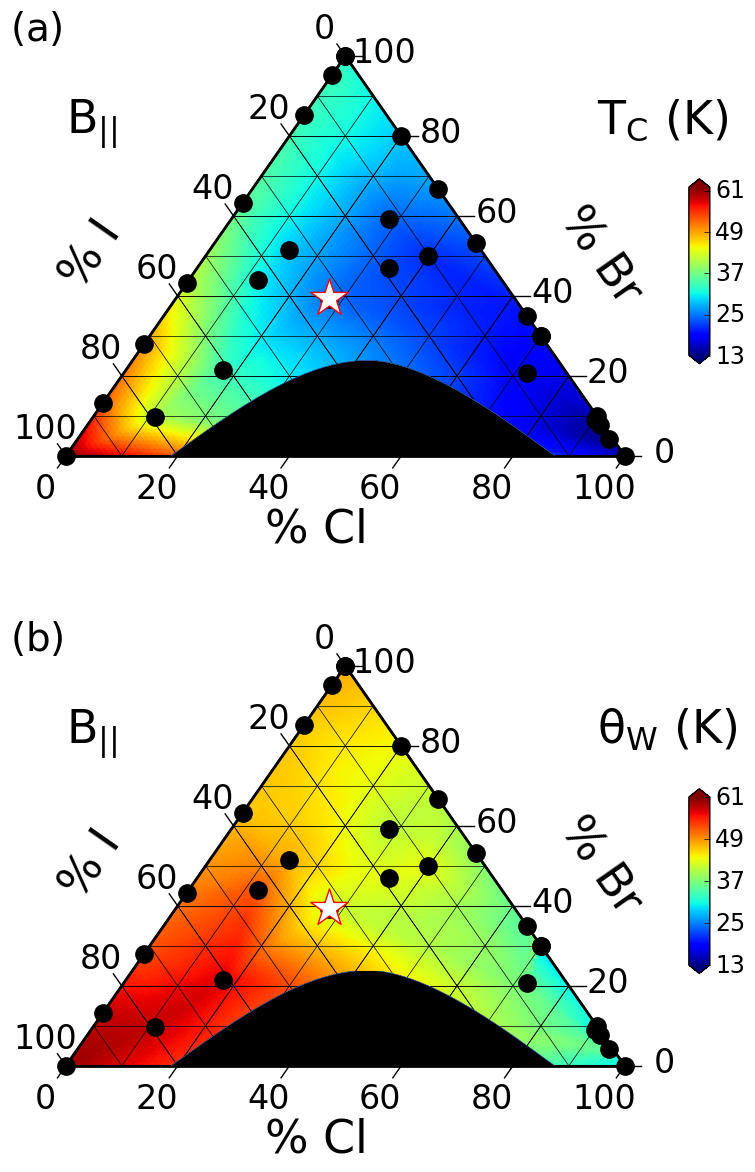}
\caption{\label{fig:PD}
(a) Triangular phase diagram of $T_C$ as a function of composition in \CBI\ with field in the plane ($B_\|$).
\textcolor{black}{The star symbol near the center is a composition with 27\% Cl (bottom axis), 40\% Br (right axis), and 33\% I (left axis), corresponding to Cr(Cl$_{0.27}$Br$_{0.40}$I$_{0.33}$)$_3$~=~CrCl$_{0.8}$Br$_{1.2}$I$_{1.0}$.}
(b) Triangular phase diagram of $\Theta_W$.
The color maps are produced by a linear interpolation between the 27 discrete data points, each represented by a black circle.
Some of the two-halide samples are the same as in ref.~\cite{abramchuk_controlling_2018}.
}
\rule{\textwidth}{0.5pt}
\end{figure}

%%%%%%%%%%%%%%%%%%%%% FIGURE 1%%%%%%%%%%%%%%%%%%%%%
Single crystals of \CBI\ were grown via a chemical vapor transport (CVT) technique from the parent compounds \CC, \CB, and \CI\ (see Fig.~\ref{fig:CBI} and Methods).
We refer to these single-phase solid-solutions as \emph{VdW alloys}.
The layered structure of the VdW alloys is illustrated in Fig.~\ref{fig:CBI}.
Each layer is a 2D honeycomb lattice made of edge-sharing octahedra around the Cr$^{3+}$ ions.
There are three electrons in the $t_{2g}$ levels of Cr$^{3+}$, giving rise to $S=3/2$ regardless of the halide ratios in \CBI\ (Fig.~\ref{fig:CBI}).
We point out that mixtures of Cl/Br (\ClBr) and Br/I (\BrI) have been reported before~\cite{abramchuk_controlling_2018}, but the three-halide series, the frustrated regime, and the bulk metamagnetic transition are presented here for the first time.
We could not grow crystals of \ClI\ due to the significant size difference between Cl and I.
A black contour in the phase diagrams of Figs.~\ref{fig:PD} and \ref{fig:F} marks the approximate region of insolubility.
Therefore, the intermediate size of Br seems to be crucial in the formation of \CBI.

%%%%%%%%%%%%%%%%%%%%% FIGURE 2%%%%%%%%%%%%%%%%%%%%%
%%%%%%%%%%%%%%%%%%%%% FIGURE 2%%%%%%%%%%%%%%%%%%%%%

%%%%%%%%%%%%%%%%%%%%%%%%%%%%%%%%%%%%%%%%%%%%%%%%%%%
\subsection*{Triangular Phase Diagrams}
%%%%%%%%%%%%%%%%%%%%%%%%%%%%%%%%%%%%%%%%%%%%%%%%%%%
The phase diagrams of Fig.~\ref{fig:PD} show that both $T_C$ and $\Theta_W$ are controlled continuously by tuning the halide composition.
This expands the list of available VdW magnets from the parent compounds, \CC, \CB, and \CI, to a continuum of compositions \CBI\ with tunable magnetic properties.
Both $T_C$ and $\Theta_W$ acquire minimum values near \CC, intermediate values near \CB, and maximum values near \CI.
We used a linear interpolation to produce the color maps in Fig.~\ref{fig:PD} based on the magnetic susceptibility data from 27 samples.
For each sample, the data were fitted to a Curie-Weiss (CW) expression, $\chi=\chi_0+C/\left(T-\Theta_W\right)$, where $\chi_0$ is a temperature-independent contribution to susceptibility, $C$ is the Curie constant related to the effective magnetic moment ($\mu=\mu_B\sqrt{8C}$), and $\Theta_W$ is the Weiss temperature related to the exchange coupling ($J=k_B\Theta_W/2S(S+1)$).

%%%%%%%%%%%%%%%%%%%%% FIGURE 3%%%%%%%%%%%%%%%%%%%%%
\begin{figure}%[htb!]
\centering
\includegraphics[width=0.45\textwidth]{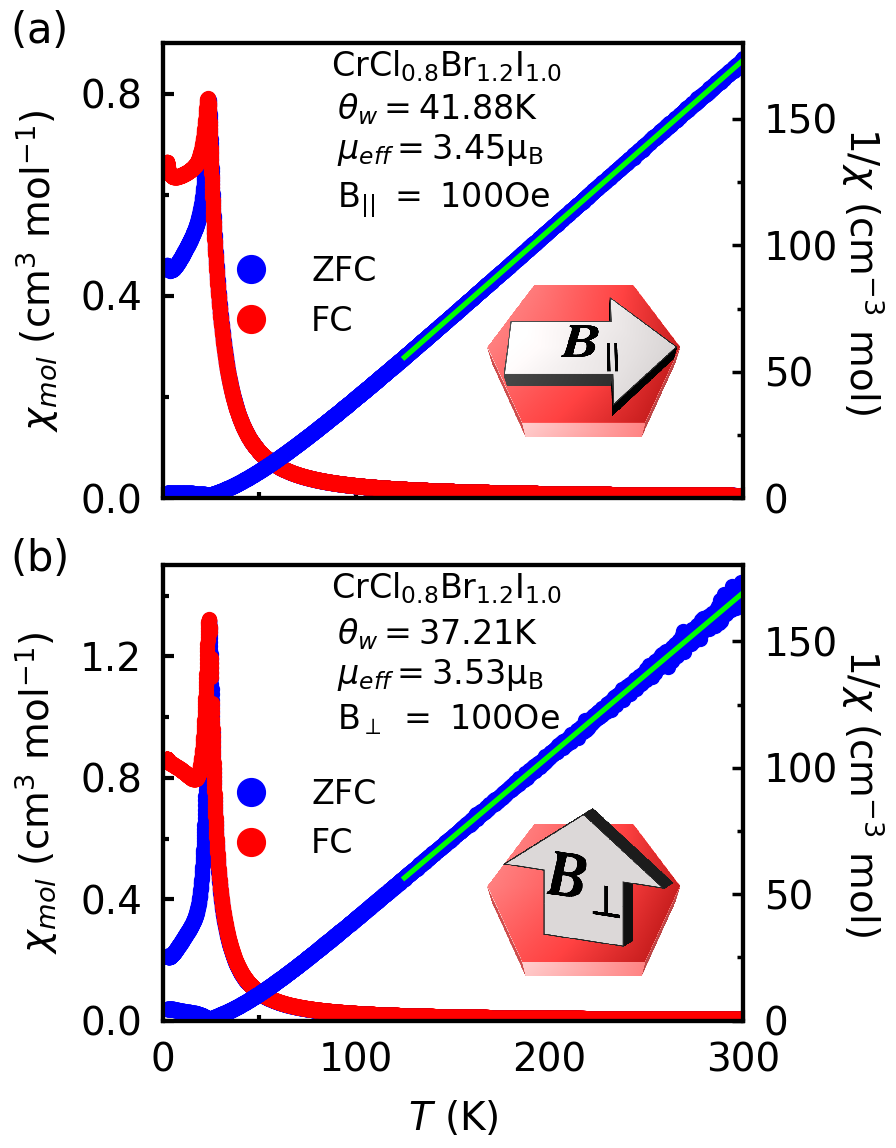}
\caption{\label{fig:CW}
Representative Curie-Weiss analyses on \CCC\ measured with field (a) parallel and (b) perpendicular to the honeycomb layers.
\textcolor{black}{The observation of a single sharp peak rules out disorder or chemical inhomogeneity (see also Fig. S3).}
Both the zero-field cooled (ZFC, blue) and field-cooled (FC, red) data are presented.
Solid green lines show the Curie-Weiss fit to the ZFC data.
The Weiss temperature $\Theta_W$ and effective moment $\mu_{\textrm{eff}}$ are comparable between the $B_\|$ and $B_{\perp}$ configurations.
\rule{\textwidth}{0.5pt}
}
\end{figure}
%%%%%%%%%%%%%%%%%%%%% FIGURE 3%%%%%%%%%%%%%%%%%%%%%

A representative CW analysis is shown in Figs.~\ref{fig:CW}(a) and (b) for \CCC, the sample marked as a star near the center of the phase diagrams in Fig.~\ref{fig:PD}.
The magnetization of all samples were measured with the magnetic field both parallel ($B_\|$) and perpendicular ($B_\perp$) to the honeycomb planes.
The effective moment evaluated by the CW analysis in all samples is close to 3.87\,$\mu_B$ corresponding to Cr$^{3+}$ (Fig.~\ref{fig:PD} and S3).
The transition temperature for each sample can be determined from the peak in either $\chi(T)$ or $d\chi(T)/dT$.
Although the results are comparable, we choose the latter criterion because it yields less uncertainty (Supplementary Material).
The triangular phase diagrams of Fig.~\ref{fig:PD} are constructed from the CW analysis on all samples in the $B_\|$ configuration.
The results for $B_{\perp}$ configuration are similar and presented in the Supplementary Material (SM).

%%%%%%%%%%%%%%%%%%%%%%%%%%%%%%%%%%%%%%%%%%%%%%%%%%%
\subsection*{Magnetic Frustration}
%%%%%%%%%%%%%%%%%%%%%%%%%%%%%%%%%%%%%%%%%%%%%%%%%%%
%%%%%%%%%%%%%%%%%%%%% FIGURE 4%%%%%%%%%%%%%%%%%%%%%
\begin{figure}%[htb!]
\centering
\includegraphics[width=0.45\textwidth]{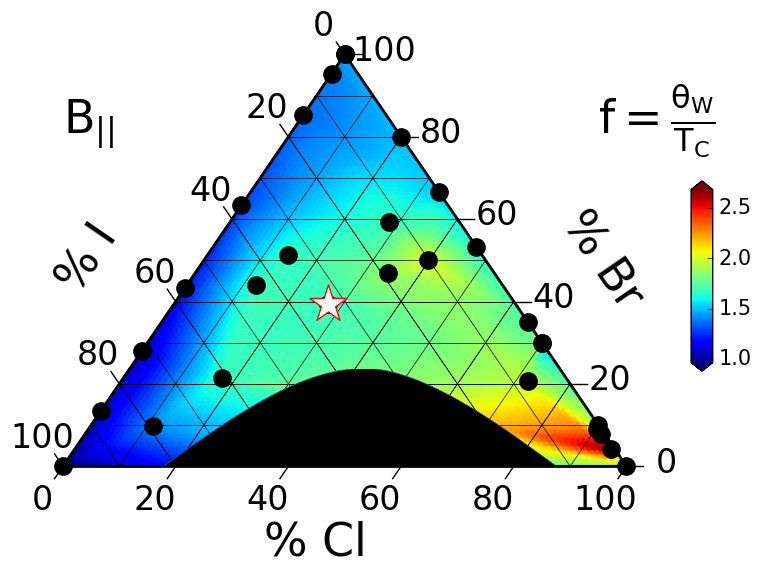}
\caption{\label{fig:F}
Frustration index ($f=\Theta_W/T_C$) as a function of composition in \CBI.
\textcolor{black}{This map is only a qualitative measure of frustration, especially since the $f$-index was originally proposed for isotropic (not anisotropic) magnets~\cite{ramirez_strongly_1994}.}
}
\rule{\textwidth}{0.5pt}
\end{figure}
%%%%%%%%%%%%%%%%%%%%% FIGURE 4%%%%%%%%%%%%%%%%%%%%%
A close inspection of the colors in Fig.~\ref{fig:PD} reveals a subtle point.
Whereas the $T_C$ and $\Theta_W$ values are nearly identical at the left corner of the phase diagram, they are quite different at the right corner.
This observation implies the presence of a moderate magnetic frustration in the compositions near \CC.
To clarify, we construct a triangular phase diagram of the frustration index, $f=\Theta_W/T_C$, in Fig.~\ref{fig:F}.
A large frustration index indicates a small $T_C$ relative to the interaction strength $J \propto \Theta_W$~\cite{ramirez_strongly_1994}.
Figure~\ref{fig:F} shows that the magnetic frustration is large near \CC, maximizes at the composition \CBmax, and gradually disappears toward either \CB\ or \CI.
Recent theoretical and experimental works have suggested that \CI\ could be a frustrated VdW material due to the Kitaev interaction~\cite{xu_interplay_2018,lee_fundamental_2020}.
However, Fig.~\ref{fig:F} suggests that compositions near \CC, specifically \CBmax, are more promising to look for the frustration effects.
This is consistent with the competition scenario between the SOC of the transition-metal and ligand (or equivalently, between D and $\lambda$ in Eq.~\ref{eq:HCOMP}), because the atomic SOC in Cr (90~meV) is comparable to Cl (40~meV), but much smaller than Br (220~meV) and I (580~meV).

%%%%%%%%%%%%%%%%%%%%%%%%%%%%%%%%%%%%%%%%%%%%%%%%%%%
\subsection*{Density Functional Theory}
%%%%%%%%%%%%%%%%%%%%%%%%%%%%%%%%%%%%%%%%%%%%%%%%%%%
%%%%%%%%%%%%%%%%%%%%% FIGURE 5%%%%%%%%%%%%%%%%%%%%%
\begin{figure}%[htb!]
\centering
\includegraphics[width=0.45\textwidth]{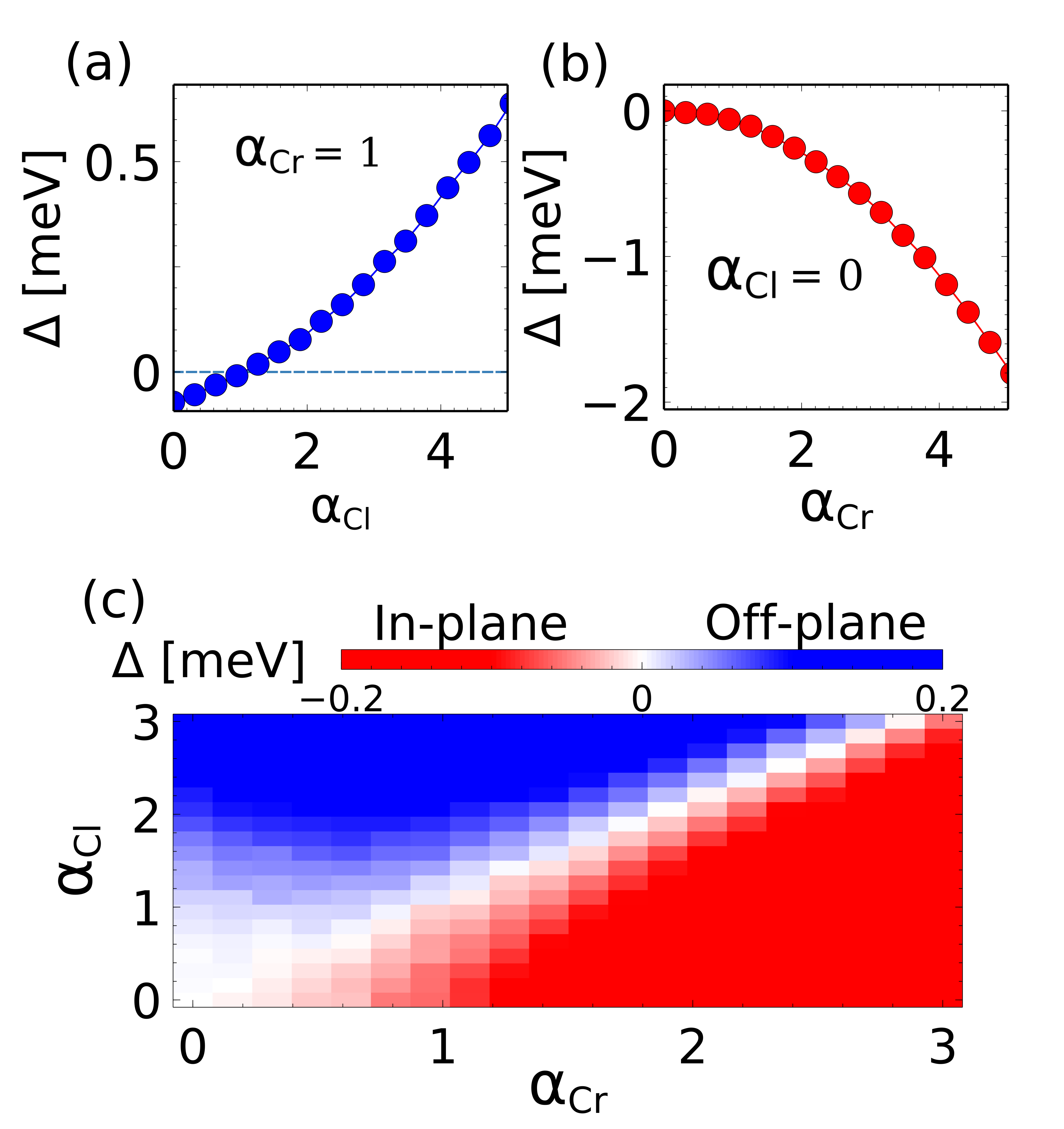}
\caption{\label{fig:SOC}
(a) The magnetic anisotropy energy $\Delta$ plotted as a function of Cl SOC ($\alpha_{\mathrm{Cl}}$) by fixing the SOC of Cr to unity ($\alpha_{\mathrm{Cr}}=1$).
(b) $\Delta$ plotted as a function of the Cr SOC by switching off the Cl SOC ($\alpha_{\mathrm{Cl}}=0$).
(c) The anisotropy energy as a function of the SOC in Cr and Cl, showing a competition between in-plane and out-of-plane anisotropies tuned by $\alpha_{\mathrm{Cr}}$ and $\alpha_{\mathrm{Cl}}$.
}
\rule{\textwidth}{0.5pt}
\end{figure}
%%%%%%%%%%%%%%%%%%%%% FIGURE 5%%%%%%%%%%%%%%%%%%%%%
The competition scenario in Eq.~\ref{eq:HCOMP} and the resulting magnetic frustration in Fig.~\ref{fig:F} can be demonstrated from first-principles.
We performed DFT calculations for the pure \CC\ by controlling the strength of the SOC contributions from Cr and Cl individually.
The DFT Hamiltonian can be written as~\cite{lado_origin_2017}
\begin{equation}
	\mathcal{H}_{DFT} = \mathcal{H}_0 + \alpha_{Cr} \mathcal{H}^{Cr}_{SOC} + \alpha_{Cl} \mathcal{H}^{Cl}_{SOC}
	\label{eq:HDFT}
\end{equation}
where $\mathcal{H}_0$ is the non-relativistic Hamiltonian in the absence of any SOC ($\alpha_{Cr}=\alpha_{Cl}=0$).
$\mathcal{H}^{Cr}_{SOC}$ is the SOC correction in Cr and $\mathcal{H}^{Cl}_{SOC}$ is the SOC correction in Cl.
By definition, the physical scenario in \CC\ corresponds to $\alpha_{Cr} = \alpha_{Cl} =1$, consistent with the atomic values of the SOC mentioned earlier~\cite{martin_table_1971}.
We can effectively tune the $\mathcal{H}_{DFT}$ from \CC\ to \CB\ and \CI\ by increasing $\alpha_{Cl}$.

Partitioning the Hamiltonian in Eq.~\ref{eq:HDFT} enables us to independently control the contribution from Cr and the halide (Cl, Br, I) to the magnetic anisotropy and trace the easy-plane versus easy-axis anisotropy.
The anisotropy energy of the system is defined as $\Delta = E_{\rightarrow} - E_{\uparrow}$, where $E_{\rightarrow}$ and $E_{\uparrow}$ are the total DFT energies of the in-plane and out-of-plane FM states, respectively.
Thus, a negative (positive) $\Delta$ corresponds to easy-plane (easy-axis).
In the following, we present the analysis in two steps.

First, we show the evolution of the anisotropy energy as a function of the halide SOC in Fig.~\ref{fig:SOC}(a) by fixing the Cr SOC to $\alpha_{Cr}=1$.
With increasing $\alpha_{Cl}$, $\Delta$ gradually approaches zero from negative and turns positive, i.e. the in-plane anisotropy is gradually replaced by out-of-plane anisotropy.
In \CC, where $\alpha_{Cr}=\alpha_{Cl}=1$, the system conserves its easy-plane anisotropy, in agreement with the experimental observation.
The alloying process effectively increases the ligand SOC ($\alpha_{Cl}>1$) and leads to $\Delta>0$, corresponding to the out-of-plane anisotropy found in CrBr$_3$ and CrI$_3$.

Second, we show in Fig.~\ref{fig:SOC}(b) the evolution of the anisotropy as a function of $\alpha_{Cr}$ by switching off the SOC of Cl ($\alpha_{Cl}=0$).
We observe a growing tendency toward in-plane anisotropy (negative $\Delta$) by increasing the Cr SOC.
This result confirms that the Cr SOC favors in-plane anisotropy and competes with the halide SOC.

The generic competition between the SOC of the magnetic atom (Cr) and the ligand (Cl) is mapped in Fig.~\ref{fig:SOC}(c), showing that the anisotropy direction is controlled by the relative strength of the two SOC constants, $\alpha_{Cr}$ and $\alpha_{Cl}$.
Notice that \CC\ is on the borderline between the in-plane and out-of-plane anisotropy.
Thus, the $D$ and $\lambda$ terms in Eq.~\ref{eq:HCOMP} nearly cancel each other and drive \CC\ toward a frustrated regime (Fig.~\ref{fig:F}).
In contrast, the strong ligand SOC in \CB\ and \CI\ induces a robust out-of-plane FM order and relieves the magnetic frustration.
\textcolor{black}{In other words, $H_{\textrm{frust.}}$ becomes perturbatively irrelevant at low energies in Eq.~\ref{eq:HCOMP} when the system has a robust easy-axis.}
We point out that a bare DFT calculation could erroneously predict an easy-axis, instead of easy-plane, in \CC~\cite{webster_strain-tunable_2018}.
We had to implement a small (2\%) lattice compression to match the DFT results with the experimental in-plane anisotropy in \CC\ (details in the SM).
We also present DFT results for \CB\ in the SM that correctly confirm out-of-plane anisotropy.

%%%%%%%%%%%%%%%%%%%%%%%%%%%%%%%%%%%%%%%%%%%%%%%%%%%
\subsection*{Metamagnetic Transition}
%%%%%%%%%%%%%%%%%%%%%%%%%%%%%%%%%%%%%%%%%%%%%%%%%%%
%%%%%%%%%%%%%%%%%%%%% FIGURE 6%%%%%%%%%%%%%%%%%%%%%
\begin{figure}%[htb!]
\centering
\includegraphics[width=0.45\textwidth]{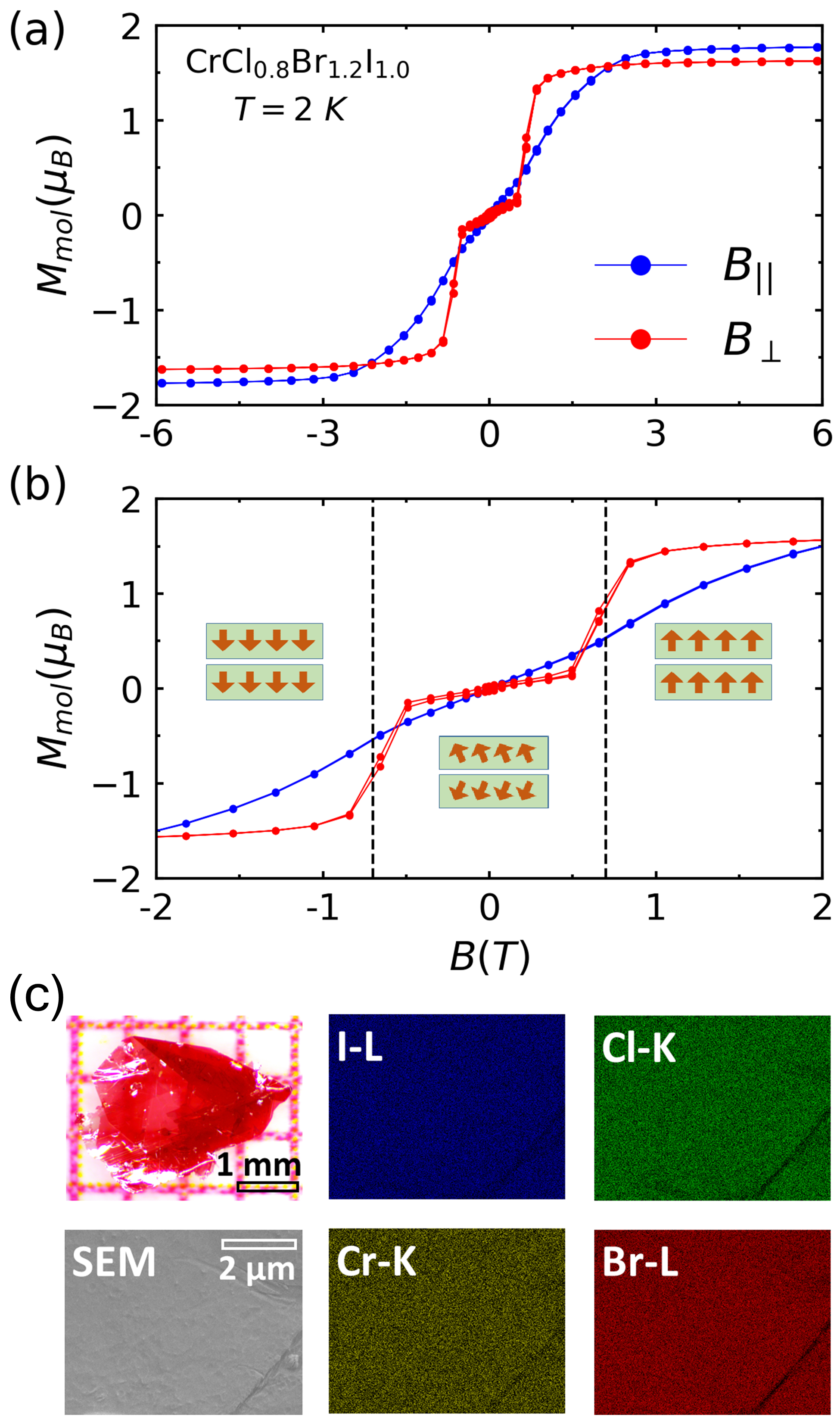}
\caption{\label{fig:MM}
(a) Magnetization plotted as a function of field in \CCC.
The red and blue data correspond to the field perpendicular ($B_\perp$) and parallel ($B_\|$) to the honeycomb layers, respectively.
The saturated moment is consistent with $S=3/2$ in Cr$^{3+}$.
(b) Magnified view of the spin-canting transition at $B_\perp=0.7$~T.
(c) Optical image, SEM image, and EDX color maps reveal a uniform distribution of Cr (yellow, $K$-edge), Cl (green, $K$-edge), Br (red, $L$-edge), and I (blue, $L$-edge) in \CCC.
}
\rule{\textwidth}{0.5pt}
\end{figure}
%%%%%%%%%%%%%%%%%%%%% FIGURE 6%%%%%%%%%%%%%%%%%%%%%
The first important finding in our experiments was a frustrated regime near \CC\ where the effects of $D$ and $\lambda$ in Eq.~\ref{eq:HCOMP} nearly cancel out, allowing $H_{\text{frust.}}$ to show its effect.
Our second exciting finding is that the inter-layer magnetic coupling in the VdW alloys is different from the parent compounds.
Recently, a metamagnetic (MM) transition has been reported at 0.7~T in the bilayers of \CI\ due to a field-induced change of interlayer coupling from AF to FM~\cite{huang_layer-dependent_2017,jiang_controlling_2018}.
The AF-FM switching of the inter-layer coupling produces a sharp step in the $M(H)$ curves of \CI\ bilayers, corresponding to a spin-flop transition.
This effect has been utilized in a spin-filter magnetic tunnel junction where a step-like magnetoresistance has been observed at the MM transition~\cite{song_giant_2018,klein_probing_2018}.
Figure~\ref{fig:MM}(a) shows a similar MM transition in the bulk crystals of \CCC\ as a function of field.
Notice that the transition is observed only when the field is perpendicular to the plane ($B_\perp$).
A magnified view of the $M(B)$ curves in Fig.~\ref{fig:MM}(b) shows that the transition occurs at $B_{\perp}=0.7$~T, the same field at which \CI\ bilayers undergo a spin-flip transition~\cite{huang_layer-dependent_2017,jiang_controlling_2018}.
However, the MM transition in \CCC\ is less sharp than in the bilayers of \CI~\cite{huang_layer-dependent_2017}.
As such, we assign this transition to a field-induced spin-canting, instead of spin-flop, as illustrated in the inset of Fig.~\ref{fig:MM}(b).

Chemical inhomogeneity or disorder do not play a role in either the MM transition or magnetic frustration.
We performed energy-dispersive X-ray spectroscopy (EDX) on each sample inside a scanning electron microscope (SEM).
The SEM-EDX color maps in Fig.~\ref{fig:MM}(c) confirm a uniform distribution of Cr, Cl, Br, and I atoms in a \CCC\ crystal and rule out a phase separation scenario.
\textcolor{black}{In materials with impurity phases or inhomogeneous distribution of elements, the color maps reveal regions of dark and light shade~\cite{abramchuk_tuning_2019}.}
A table of EDX results is reported in the SM.
Furthermore, all three-halide VdW alloys show a single sharp magnetic transition (Figs.~\ref{fig:CW}, S2, and S3), similar to the parent compounds~\cite{mcguire_crystal_2017}.
\textcolor{black}{If the samples were disordered or chemically inhomogeneous, we would have expected either a rounded transition or multiple transitions, which is clearly not the case.}

%%%%%%%%%%%%%%%%%%%%%%%%%%%%%%%%%%%%%%%%%%%%%%%%%%%
\section*{\label{sec:conclusions}Summary and Outlook}
%%%%%%%%%%%%%%%%%%%%%%%%%%%%%%%%%%%%%%%%%%%%%%%%%%%
% To summarize, the \CBI\ series can be regarded as VdW alloys where the magnetic properties are controlled by tuning the ratio of the halides.
% %
% The fundamental parameter that is being tuned is the SOC of the ligand atoms.
% %
Despite the conventional wisdom to tune magnetism by doping on the cation site where the magnetic moment resides~\cite{bi_room-temperature_2018,zarick_mixed_2018}, we highlight the remarkable effect of mixing the anions to synthesize a continuum of VdW magnets with chemical control over the ligand SOC.
\textcolor{black}{Note that the change of SOC from Cl to I is enormous (more than one order of magnitude) based on the atomic values of SOC~\cite{martin_table_1971}, whereas the change of local geometry and bond angle is minimal.
For example, the Cr-ligand-Cr bond angles are 93.9$^\circ$ and 93.3$^\circ$ in the low-temperature structure of \CC~and \CI, respectively~\cite{morosin_xray_1964,mcguire_coupling_2015}.
Thus, the dominant tuning parameter in our phase diagrams is the ligand SOC.}

The chromium trihalides provided a unique opportunity for our experiments due to their remarkable chemical tunability.
Because the SOC of Cl$<$Cr$<$Br$<$I, we were able to tune the SOC of the magnetic ion and the ligands against each other.
Without making a solid-solution of three halides and mapping the phase diagram of \CBI, it would have been impossible to find the frustrated regime and the change of inter-layer coupling.

\textcolor{black}{We point out that the alloying process inevitably leads to bond randomness.
Based on the remarkable agreement between our mean-field DFT approach and the experimental results, it appears that the bond randomness at an atomic level does not affect the magnetic anisotropy and ordering dramatically.
However, it would be an interesting future direction to go beyond the mean-field level and explore the role of bond randomness in VdW alloys near the Kitaev limit~\cite{knolle_bond-disordered_2019}.
}

The methods presented here can, in principle, be extended to chalcogenides.
Examples of chalcogen-based VdW magnets are the itinerant ferromagnet \FGT~\cite{fei_two-dimensional_2018,deng_gate-tunable_2018}, the FM insulator \CGT~\cite{gong_discovery_2017}, and the AFM layered compound FePS$_3$~\cite{lee_ising-type_2016,wang_raman_2016}.
One advantage of the chalcogenides is their relatively high transition temperatures.
For example, $T_C=220$~K in \FGT~\cite{kim_large_2018}, $T_C=60$~K in \CGT~\cite{ji_ferromagnetic_2013}, and $T_N=118$~K in FePS$_3$~\cite{lee_ising-type_2016}.
The magnetic properties of these materials can be controlled by mixing the chalcogens S, Se, and Te.
Such efforts will expand the list of desirable VdW magnets~\cite{gong_two-dimensional_2019,burch_magnetism_2018}.
But more importantly, they will lead to new regimes (such as frustration) and new phenomena (such as spin-canting transition).

As mentioned earlier, the magnetic ordering in \CI\ is known to acquire a layer-dependent structure in the atomically thin limit (single layer, bilayer, and trilayer)~\cite{huang_electrical_2018,jiang_controlling_2018,wang_electric-field_2018,seyler_ligand-field_2018,klein_probing_2018,chen_direct_2019}.
\textcolor{black}{An important question is whether the MM transition at 0.7~T in the bulk crystals of \CCC\ is due to a change of stacking sequence or a structural transition at low temperatures.
This question could be answered by low-temperature synchrotron X-ray experiments in the future.}

In the SM, we show that the alloyed crystals are as exfoliable as their parent structures.
It will be interesting to exfoliate the bulk crystals of \CCC\ and isolate bilayers and trilayers to look for a richer pattern of MM transitions by measuring the magneto-optical Kerr Effect (MOKE).
Such heterostructures will be an invaluable resource for new devices, such as spin-filter tunneling junctions~\cite{song_giant_2018,klein_probing_2018,wang_very_2018,kim_one_2018} and magnon-assisted tunneling devices~\cite{ghazaryan_magnon-assisted_2018}.
%

%%%%%%%%%%%%%%%%%%%%%%%%%%%%%%%%%%%%%%%%%%%%%%%%%%%%%%%%%%%%%%%%%%%%%%%%%%%%%%%%%%%%%%%%%%%%%%%%%%%%%%%%%%%%%%%%%%%%%%%%%%%
%%%%%%%%%%%%%%%%%%%%%%%%%%%%%%%%%%%%%%%%%%%%%%%%%%%%%%%%%% METHODS %%%%%%%%%%%%%%%%%%%%%%%%%%%%%%%%%%%%%%%%%%%%%%%%%%%%%%%%
%%%%%%%%%%%%%%%%%%%%%%%%%%%%%%%%%%%%%%%%%%%%%%%%%%%%%%%%%%%%%%%%%%%%%%%%%%%%%%%%%%%%%%%%%%%%%%%%%%%%%%%%%%%%%%%%%%%%%%%%%%%
\section*{Materials and Methods}
% The materials and methods section should provide sufficient information to allow replication of the results. Begin with a section titled Experimental Design describing the objectives and design of the study as well as pre-specified components.
%
% In addition, include a section titled Statistical Analysis at the end that fully describes the statistical methods with enough detail to enable a knowledgeable reader with access to the original data to verify the results. The values for N, P, and the specific statistical test performed for each experiment should be included in the appropriate figure legend or main text.

\subsection*{Crystal Growth}
The crystals of \CBI\ were grown by the CVT method from \CC, \CB, and \CI\ precursors mixed with appropriate mole ratios (see Table S1 in the SM).
In a typical CVT cycle, 400~mg of the starting mixture was placed inside a 6$^{\prime\prime}$-long fused silica tube and held at 650~\C, with a 200~\C\ temperature gradient, for 72~h.
\CC\ crystals were grown by vacuum sublimation of polycrystalline \CC\ at 650~\C\ for 72~h.
\CB\ crystals were grown by annealing a mixture of Cr+0.75TeBr$_4$ at 700~\C\ for 72~h.
\CI\ crystals were grown by annealing a mixture Cr+1.5I$_2$ at 650~\C\ for 72~h.
The starting chemicals, Cr(99.99\%), \CC\ (99.9\%), TeBr$_4$ (99.9\%), and I$_2$ (99.8\%) were purchased from Alfa Aesar.
The crystals grew in the form of thin plates, several millimeters across and less than 1~$\mu$m thick.

\subsection*{Electron Microscopy}
The morphology and chemical composition of crystals were studied by scanning electron microscopy (SEM) and energy dispersive X-ray spectroscopy (EDX), using a JEOL 7900F field-emission scanning electron microscope (FESEM) equipped with an EDAX detector.

\subsection*{Magnetic Measurements}
Magnetic measurements were performed using a vibrating sample magnetometer (VSM) inside a Quantum Design MPMS-3.
Thin plate-like crystals were mounted on a low-background quartz or sapphire holder for the $B_\|$ and $B_\perp$ measurements, respectively.

\subsection*{Density Functional Theory}
Density functional theory (DFT) calculations were performed using the all-electron Elk code~\cite{noauthor_elk_nodate}.
Lattice relaxations were performed with Quantum Espresso~\cite{giannozzi_quantum_2009}, PBEsol functional, and PAW pseudopotentials.
A $2\%$ lattice compression was imposed to match the DFT results with the experimental observations (see SM for details)~\cite{webster_strain-tunable_2018,bruno_spin-wave_1991}.
The total energy convergence was set to 0.027~$\mu$eV to account for the small energy differences involved in magnetic anisotropies.

%%%%%%%%%%%%%%%%%%%%%%%%%%%%%%%%%%%%%%%%%%%%%%%%%%%%%%%%%%%%%%%%%%%%%%%%%%%%%%%%%%%%%%%%%%%%%%%%%%%%%%%%%%%%%%%%%%%%%%%%%%%
%%%%%%%%%%%%%%%%%%%%%%%%%%%%%%%%%%%%%%%%%%%%%%%%%%%%%%%%% REFERENCE %%%%%%%%%%%%%%%%%%%%%%%%%%%%%%%%%%%%%%%%%%%%%%%%%%%%%%%
%%%%%%%%%%%%%%%%%%%%%%%%%%%%%%%%%%%%%%%%%%%%%%%%%%%%%%%%%%%%%%%%%%%%%%%%%%%%%%%%%%%%%%%%%%%%%%%%%%%%%%%%%%%%%%%%%%%%%%%%%%%

% Your references go at the end of the main text, and before the
% figures.  For this document we've used BibTeX, the .bib file
% scibib.bib, and the .bst file Science.bst.  The package scicite.sty
% was included to format the reference numbers according to *Science*
% style.

\bibliography{sciadvbib}
\bibliographystyle{ScienceAdvances}

\pagebreak

\noindent \textbf{Acknowledgements:}
% Acknowledgments should be gathered into a paragraph after the final numbered reference. This section should also include
% * complete funding information,
% * a description of each authorÕs contribution to the paper,
% * a listing of any competing interests of any of the authors (all authors must also fill out the Conflict of Interest form), and,
% * a section on data and materials availability, information about the location of the data if not included in the paper, including **accession numbers** to any data relating to the paper and deposited in a public database.
%
F.T. acknowledges support from the National Science Foundation under Grant No. DMR-1708929.
J.Y.C acknowledges support from the National Science Foundation under Grant No. DMR-1700030.
Y.R. acknowledges support from the National Science Foundation under Grant No. DMR-1712128.
J.L.L. acknowledges the computational resources provided by the Aalto Science-IT project.
M.D and K.S.B. acknowledge support from the US Department of Energy (DOE), Office of Science, Office of Basic Energy Sciences under award no. DE-SC0018675.
\\
\noindent \textbf{Author Contributions:} TT, JNT, and MA grew the crystals and performed measurements. TT, JNT, and FB analyzed data. JLL performed DFT calculations. GTM and JYC characterized the samples. MD and KSB performed exfoliation. All authors contributed to writing the manuscript. YR and FT initiated the project.\\
\noindent \textbf{Competing Interests:} The authors declare that they have no competing financial interests.\\
\noindent \textbf{Data and materials availability:} Additional data and materials are available online.\\
\noindent\textbf{Supplementary Material:} This paper has supplementary material in a separate PDF file.

% For your review copy (i.e., the file you initially send in for
% evaluation), you can use the {figure} environment and the
% \includegraphics command to stream your figures into the text, placing
% all figures at the end.  For the final, revised manuscript for
% acceptance and production, however, PostScript or other graphics
% should not be streamed into your compliled file.  Instead, set
% captions as simple paragraphs (with a \noindent tag), setting them
% off from the rest of the text with a \clearpage as shown  below, and
% submit figures as separate files according to the Art Department's
% instructions.

% \clearpage

% \noindent {\bf Fig. 1.} Please do not use figure environments to set
% up your figures in the final (post-peer-review) draft, do not include graphics in your
% source code, and do not cite figures in the text using \LaTeX\
% \verb+\ref+ commands.  Instead, simply refer to the figure numbers in
% the text per {\it Science\/} style, and include the list of captions at
% the end of the document, coded as ordinary paragraphs as shown in the
% \texttt{sciadvfile.tex} template file.  Your actual figure files should
% be submitted separately.

\end{document}